# Hydrogenation of graphene in view of odd electrons correlation

*E. F. Sheka, N. A. Popova*
*Peoples' Friendship University of Russia*
*Miklukho-Maklaya 6, Moscow 117198, Russia*
E-mail: sheka@icp.ac.ru

**Abstract.** The paper presents evidence of a rather strong correlation of odd electrons in the singlet state of graphene. Due to the correlation, the chemical modification of graphene can be considered following a certain algorithmic computational procedure. Originated due to the correlation and distributed over the carbon atoms of graphene membrane with fraction numbers $N_{DA}$, effectively unpaired electrons lay the algorithm foundation. The highest $N_{DA}$ value points to the target atom that enters a chemical reaction at the considered step. Following the pointers, a stepwise design of polyderivatives can be performed. Applied to the hydrogenation, the algorithmic design has exhibited that graphene hydrogenation should be attributed to a highly complicated event, whose final hydride products depend on a number of factors such as: 1) the manner of the graphene membrane fixation; 2) the accessibility of the membrane both sides to hydrogen; 3) the composition (molecular or atomic) of the hydrogen. In general, the hydride formation is multimode in regards composition and structure. Thus, the formation of 100% hydride with regular chairlike hexagonal packing of CH units which can be attributed to graphane is possible if only the graphene membrane is fixed over perimeter while its basal plane is accessible to hydrogen atoms from both sides.

## 1. Odd electrons correlation and chemical modification of graphene

Just recently basic grounds have been considered of a 'breakthrough' nanotechnology aimed at the production and safe storage of solid molecular hydrogen of both high density and purity "…via intercalation into multilayer graphane-like (carbohydride) nanostructures" [1, 2]. The studies make the understanding of the graphene interaction with both molecular and atomic hydrogen extremely actual. To proceed with the technology, one should answer two key questions:
1)  What are 'graphane-like (carbohydride)' nanostructures?
2)  How do the structures provide the formation and storage of solid molecular hydrogen?

The first question directly addresses chemical modification of either graphite or graphene caused by the hydrogen adsorption on the relevant surfaces. The second question concerns the intermolecular interaction between the modified graphene layers involving hydrogen. In the current paper we shall restrict ourselves by looking for the answer to the first question.

Experimentally has been known for a long time that molecular adsorption of hydrogen on graphite is extremely weak and that its hydrogenation occurs in plasma containing atomic hydrogen only. Owing to this, the interest to the study of the graphite hydrogenation was steadily lowering until the appearance of graphene has drastically activated the interest so that the hydrogenation of graphene has become a problem of urgent interest. The widely known Science-2009 paper [3] has heralded the discovery of a new two-dimensional crystalline material that was the product of the graphene hydrogenation in the hydrogen plasma and that was named as graphane. The discovery has stimulated a dense stream of computational studies devoted to the interaction of graphene with hydrogen as well as to the effect of the hydrogenation on fundamental properties of graphene [4-31], but a few. However, the majority of quantum computationists involved in the studies have considered the motion of weakly bound odd electrons of graphene with different spins not correlated. Only this fact can justify them using computational techniques based on single restricted closed-shell determinants, mainly in a manner of various DFT schemes. Sometimes, these DFT computational techniques have even been heralded as indisputable standard of the computations in the field. However, the electron behaviour in graphene is much richer that will be shown in the current paper by the example of the graphene hydrogenation.

The electron correlation is evidently a prerogative of quantum molecular theory. As for graphene, in spite of formally monatomic crystalline structure, its properties are evidently governed by the behaviour of odd electrons of hexagonal benzenoid units. The only thing that we know for sure is that the interaction between these electrons is weak. But how weak is it? Is it, nevertheless, enough to provide a tight covalent pairing when two electrons with different spins occupy the same place in space or, oppositely, is it too weak for this and the two electrons are located in different spaces thus becoming spin correlated? This very important molecular aspect of graphene can be visualised on the platform of molecular quantum theory. And only within this theory one can get answer if the odd electrons of graphene are 'yes/no' spin correlated and what consequence of the correlation in regards basic properties of graphene can be expected.

When speaking about electron correlation, one must address the problem to the configuration interaction (CI). However, neither full CI nor any its truncated version, clear and transparent conceptually, can be applied for computations, valuable for graphene nanoscience that requires a great number of computations to be performed as well as a great number atoms to be considered [32, 33]. Owing to this, techniques based on single unrestricted open shell determinants become the only alternative. Unrestricted DFT (spin polarized, UDFT) and unrestricted Hartree-Fock (UHF) approaches form the techniques ground and are both sensitive to the electron correlation, but differently due to different dependence of their algorithms on electron spins [34, 35].

In the framework of the single-determinant approach, one can suggest three criteria that may both evidence the electron correlation and characterize its extent. The relevant characteristic parameters are the following:

1. Misalignment of energy

$$\Delta E^{RU} = E^R - E^U \geq 0 \,. \tag{1}$$

Here $E^R$ and $E^U$ present total energies calculated by using restricted and unrestricted versions of the same program;

2. The number of effectively unpaired electrons

$$N_D = trD(r|r') \neq 0 \quad \text{and} \quad N_D = \sum_A D_A, \qquad (2)$$

where $D(r|r')$ [36] and $D_A$ [37] present the total and atom-fractioned spin density caused by the spin asymmetry due to the location of electrons with different spins in different spaces;

3. Misalignment of squared spin

$$\Delta \hat{S}^2 = \hat{S}_U^2 - S(S+1) \geq 0. \qquad (3)$$

Here $\hat{S}_U^2$ presents the squared spin value calculated within the applied unrestricted technique while $S(S+1)$ presents the exact value of $\hat{S}^2$.

Criterion 1 follows from a well known fact that the electron correlation lowers the total energy [38]. Criterion 2 highlights the fact that the electron correlation is accompanied with the appearance of effectively unpaired electrons that provide the molecule radicalization [36, 37, 39]. The total number of effectively unpaired electrons depends on interatomic distance when the latter exceeds a critical value $R_{cov}^{crit}$, under which two electrons are covalently bound but above which the two electrons become unpaired, the more, the larger is the interatomic spacing. In the case of the $sp^2$ C-C bonds, $R_{cov}^{crit}$=1.395Å [40]. Criterion 3 is the manifestation of the spin contamination of unrestricted single-determinant solutions [37, 39]; the stronger electron correlation, the bigger spin contamination of the studied spin state.

Table 1 presents sets of the three parameters evaluated for a number of right-angled fragments ($n_a,n_z$) of graphene ($n_a$ and $n_z$ count the numbers of benzenoid units along armchair and zigzag edges of the fragment, respectively [41]) by using AM1 version of semiempirical UHF approach implemented in the CLUSTER-Z1 codes [42]. To our knowledge, only this program allows for computing all the above three parameters simultaneously. As seen in the table, the first two of them are definitely not zero by value, besides greatly depending on the fragment size. At the same time, their relative values are constant within ~10% accuracy pointing to a stable and size independent effect caused by the electron correlation. The attention should be called to rather big $N_D$ values, both absolute and relative. The finding evidences that the length of C-C bonds in the considered fragments exceed the critical value $R_{cov}^{crit}$=1.395Å. It should be added as well that the relation $N_D = 2\hat{S}_U^2$, which is characteristic for spin contaminated solutions in the singlet state [38], is rigidly kept over all the fragments.

The data convincingly evidence that the electron correlation in graphene is significant. In view of the correlation, peculiarities of the graphene chemistry, magnetism, and mechanics can be exhibited at the quantitative level, much as this has been done for fullerenes [33]. Similarly to the latter, chemical reactivity of graphene is atomically mapped [43-45] due to the difference in values $D_A$ in (2). The latter as well as $N_D$, properly adapted to the *NDDO* UHF computational algorithm of CLUSTER-Z1 codes [46], has the form

$$N_D = \sum_{i,j=1}^{NORBS} D_{ij} = \sum_A N_{DA} \qquad (4)$$

and

$$N_{DA} = \sum_{i \in A} \sum_{B=1}^{NAT} \sum_{j \in B} D_{ij} \ . \tag{5}$$

Here $D_{ij}$ are elements of spin density matrix that presents a measure of the electron correlation [36, 37, 39]. In the singlet state, the $N_{DA}$ values are identical to the atom free valences [37] and thus exhibit the atomic chemical susceptibility (ACS) [47]. The $N_{DA}$ distribution over atoms plots a 'chemical portrait' of the studied molecule, whose analysis allows for making a definite choice of the target atom with the highest $N_{DA}$ value to be subordinated to chemical attack by an external addend. A typical chemical portrait of a graphene fragment highlights edge atoms as those with the highest chemical activities, besides quite irregular, while exhibiting additionally the basal atoms ACS comparable with that of fullerene [43-45].

    This circumstance is the main consequence of the odd electron correlation in graphene in regards its chemical modification. Ignoring the correlation has resulted in a common conclusion about chemical inertness of the graphene atoms with the only exclusion concerning edge atoms due to which a computationists does not know the place of the first as well as consequent chemical attacks to be possible on the basal plane and has to perform calculations sorting them out over the atoms by using the lowest-total-energy (LTE) criterion (see, for example, [25]). In contrast, basing of the $N_{DA}$ value as a quantitative pointer of the target atom at any step of the chemical attack, one can suggest the algorithmic 'computational synthesis' of the molecule derivatives. Following this algorithmic procedure, a stepwise hydrogenation of graphene has been performed in the current study.

## 2. Algorithmic computational design of graphene polyhydride (CH)$_n$

Due to a vast number of computations to be performed, a (5,5) rectangular nanographene with 88 odd electrons has been chosen for the study. As seen in Table 1, its identifying parameters are within the average quantities so that the extent of the odd electrons correlation in the fragment is similar to those characteristic for large-size ones. Calculations were carried by using UHF approach implemented in semiempirical AM1 version of the CLUSTER-Z1 codes [42] whose algorithm is described in details in [48]. Absolutely perfect parameterization of the input data for carbon atoms provides a high accuracy of the results obtained, which has been repeatedly proven in numerous cases [33].

    Equilibrium structure of the fragment alongside with its ACS map, which presents the distribution of atomically-matched effectively unpaired electrons $N_{DA}$ over the fragment atoms, is shown in Fig.1. Panel *b* exhibits the ACS distribution attributed to the atoms positions thus presenting the 'chemical portrait' of the fragment. Different ACS values are plotted in different coloring according to the attached scale. The absolute ACS values are shown in panel *c* according to the atom numbering in the output file. As seen in the figure, 22 edge atoms involving 2x5 *zg* and 2x6 *ach* ones have the highest ACS thus marking the perimeter as the most active chemical space of the fragment. These atoms are highlighted by the fact that each of them posses two odd electrons, the interaction between which is obviously weaker than that for the basal atoms. Providing the latter, the electron correlation

and the extent of the electron unpairing are the highest for these atoms, therewith bigger for *zg* edges than for *ach* ones.

The hydrogenation of the fragment will start on atom 14 (star-marked in Fig.1c) according to the highest ACS in the output file. The next step of the reaction involves the atom from the edge set as well and this is continuing until all the edge atoms are saturated by a pair of hydrogen atoms each since all 44 steps are accompanied with the high-rank ACS list where edge atoms take the first place. Thus obtained hydrogen-framed graphene molecule is shown in Fig. 2 alongside with the corresponding ACS map. Two equilibrium structures are presented. The structure in panel *a* corresponds to the optimization of the molecule structure without any restriction. Once the positions of edge carbon atoms and framing hydrogen atoms were fixed, the optimization procedure was repeated leading to the structure shown in panel *c*. In what follows, we shall refer to the two structures as free standing and fixed membranes, respectively. Blue atoms in Fig. 2c alongside with framing hydrogens are excluded from the forthcoming optimization under all steps of the further hydrogenation.

Chemical portraits of the structures shown in Fig. 2b and Fig. 2d are quite similar and reveal the transformation of brightly shining edge atoms in Fig. 1b into dark spots. Actually, the addition of two hydrogen atoms to each of the edge ones saturates the valence of the latter completely, which results in zeroing ACS values, as is clearly seen in Fig. 2e. The chemical activity is shifted to the neighboring inner atoms and retains higher in the vicinity of *zg* edges, however, differently in the two cases. The difference is caused by the redistribution of C-C bond lengths of free standing membrane when it is fixed over perimeter, thus providing different starting conditions for the hydrogenation of the two membranes.

Besides the two types of initial membranes, the hydrogenation will obviously depend on 1) the type of hydrogen species in use and 2) the accessibility of the membranes sides to the hydrogen. Even these circumstances evidence the hydrogenation of graphene to be a complicated chemical event that strongly depends on the initial conditions, once divided into 8 adsorption modes in regards atomic or molecular adsorption; one- or two-side accessibility of membranes; and free or fixed state of the membranes perimeter. Six of these modes (four for atomic adsorption and two for molecular one) were investigated in the current study. Only two of the latter correspond to the experimental observation of hydrogenated specimens discussed in [3], namely: two-side and one-side atomic hydrogen adsorption on the fixed membrane. Only these two modes will be described in details in what follows. The remainder modes will be reviewed only briefly.

### 3. Two-side atomic adsorption of hydrogen on fixed membrane

The hydrogenation concerns the basal plane of the fixed hydrogen-framed membrane shown in Fig. 2c that is accessible to hydrogen atoms from both sides. The obtained products form hydrides family 1. To facilitate the further presentation of the equilibrium structures, framing hydrogen atoms will not be shown. As seen in Fig. 2e, the first hydrogenation step should occur on basal atom 13 marked by star. Since the membrane is accessible to hydrogen from both sides, we have to check which deposition of the hydrogen atom, namely, above the carbon plane ('up') or below it ('down') satisfies the LTE criterion. As seen from Chart 1, the up position is somewhat preferential and the obtained equilibrium structure H1 is shown in Fig. 3. The atomic structure is accompanied with the ACS map that makes it possible to trace the transformation of the chemical activity of the graphene fragment caused by the first deposition. Rows HK (N) in Chart 1 display intermediate graphene hydrides involving K hydrogen atoms adsorbed on the basal plane while H0 is related to (5, 5) nanographene with 44 framing hydrogen atoms. N points the number of basal carbon atom in the output file (see

Fig. 2e) to which the K$^{th}$ hydrogen atom is attached. Columns $N_{DA}$ and $N_{at}$ present the high rank ACS values and the number of atoms to which they belong. The calculated heat of formation ΔH is subjected to the LTE criterion for selecting the best isomorphs that are shown in the chart by light blue shading.

After deposition of hydrogen 1 on basal atom 13, the ACS map has revealed carbon atom 46 for the next deposition (see H1 ACS map in Fig. 3). The LTE criterion favors the down position for the second hydrogen on this atom so that we obtain structure H2 shown in Fig. 3. The second atom deposition highlights next targeting carbon atom 3 (see ACS map of H2 hydride), the third adsorbed hydrogen atom activates target atom 60, the fourth does the same for atom 17, and so forth. Checking up and down depositions in view of the LTE criterion, a choice of the best configuration can be performed and the corresponding equilibrium structures for a selected set of hydrides from H1 to H11 are shown in Fig. 3. As follows from the results, the first 8 hydrogen atoms are deposited on substrate atoms characterized by the largest ACS peaks in Fig. 2b. After saturation of these most active sites, the hydrogen adsorption starts to fill the inner part of the basal plane in a rather non-regular way therewith, which can be traced in Fig. 3 and Fig. 4. And the first hexagon unit with the cyclohexane chairlike motive (cyclohexanoid chair) is formed when the number of hydrogen adsorbates achieves 38. This finding well correlates with experimental observation of a disordered, seemingly occasionally distributed, adsorbed hydrogen atoms on the graphene membrane at similar covering [49]. A complete computational procedure is discussed in details elsewhere [50].

The structure obtained at the end of the 44$^{th}$ step is shown at the bottom of Fig. 4. It is perfectly regular, including framing hydrogen atoms thus presenting a computationally synthesized chairlike (5,5) nanographane in full accordance with the experimental observations [3].

### 4. One-side atomic adsorption of hydrogen on fixed membrane

Coming back to the first step of the hydrogenation, which was considered in the previous Section, we proceed further with the second and all the next steps of the up deposition only. As previously, the choice of the target atom at each step is governed by high-rank $N_{DA}$ values. Figure 5 present a sketch of equilibrium adsorption configurations. As seen in the figure, the sequence of target atoms repeats that one related to the two-side adsorption up to the 10$^{th}$ (actually, to the 11$^{th}$) step, after which the order of target atoms differs from that of the previous case. Starting from the 24$^{th}$ step, a part of the carbon carcass becomes concave and the carcass as a whole transforms into a canopy by the final 44$^{th}$ step. However, after a successful adsorption of the 43$^{rd}$ atom, the deposition of the 44$^{th}$ hydrogen not only turns out to be impossible but stimulates desorption of previously adsorbed atom situated in the vicinity of the last target carbon atom. As a consequence, the two hydrogens are coupled and a hydrogen molecule desorbs from the fragment. A peculiar canopy shape of the carbon carcass of hydride H44 is solely by the formation of the table-like cyclohexanoid units. However, the unit packing is only quasi-regular that may explain the amorphous character of the hydrides formed at the outer surface of graphene ripples observed experimentally [3]. A complete set of the obtained products form hydrides family 2.

As for the hydrogen coverage, Fig. 6 presents the distribution of C-H bond lengths of hydrides H44 of families 1 and 2. In both cases the distribution consists of two parts, the first of each covers 44 bonds formed with edge carbon atoms. Obviously, this part is identical for both hydrides since the bonds are related to framing atoms excluded from the optimization.

The second part covers C-H bonds formed by hydrogen atoms attached to the basal plane. As seen in the figure, in the case of hydride 1, C-H bonds are practically identical with the average length of 1.126Å and only slightly deviate from those related to framing atoms. This is just a reflection of the regular graphane-like structure of the hydride shown in Fig.4. In contrast, C-H bonds on a canopy-like carbon carcass of hydride 2 are much longer than those in the framing zone, significantly oscillate around the average value of 1.180Å and even achieve value of 1.275Å. In spite of the values greatly exceed a 'standard' C-H bond length of 1.11Å typical for benzene, those are still related to chemical C-H bonds, whilst stretched, since the C-H bond rupture occurs at the C-H distance of 1.72Å [51]. A remarkable stretching of the bonds points to a considerable weakening of the C-H interaction for hydrides 2, which is supported by the energetic characteristics of the hydrides as well.

## 5. Energetic characteristics accompanying the nanographene hydrogenation

Total coupling energy that may characterize the molecule hydrogenation can be presented as

$$E_{cpl}^{tot}(n) = \Delta H_{nHgr} - \Delta H_{gr} - n\Delta H_{at}. \qquad (6)$$

Here $\Delta H_{nHgr}$, $\Delta H_{gr}$, and $\Delta H_{at}$ are heats of formation of graphene hydride with $n$ hydrogen atoms, a pristine nanographene, and hydrogen atom, respectively. Since we are mainly interested in the adsorption on basal plane, it is worthwhile to disclose the coupling energy related to the basal adsorption in the form

$$E_{cpl}^{tot\ bs}(k) = E_{cpl}^{tot}(k+44) - E_{cpl}^{tot\ fr}(44), \qquad (7)$$

where $k = n - 44$ (K in Chart 1) numbers hydrogen atoms deposited on the basal plane and $E_{cpl}^{tot\ bs}(k)$ presents the coupling energy counted off the energy of the framed membrane $E_{cpl}^{tot\ fr}(44)$.

The tempo of hydrogenation may be characterized by the coupling energy needed for the addition of each next hydrogen atom. Attributing the energy to the adsorption on the basal plane, the perstep energy can be determined as

$$E_{cpl}^{step\ bs}(k) = E_{cpl}^{tot}(k+44) - E_{cpl}^{tot}[(k+44)-1]. \qquad (8)$$

Evidently, two main contributions, namely, the deformation of the fragment carbon skeleton (def) and the covalent coupling of hydrogen atoms with the substrate resulted in the formation of C-H bonds (cov) determine both total and perstep coupling energies. Supposing that the relevant contributions can be summed up, one may evaluate them separately. Thus, the total deformation energy can be determined as the difference

$$E_{def}^{tot}(n) = \Delta H_{nHgr}^{sk} - \Delta H_{gr}. \qquad (9)$$

Here $\Delta H_{nHgr}^{sk}$ presents the heat of formation of the carbon skeleton of the *n*-th hydride at the *n*-th stage of hydrogenation and $\Delta H_{gr}$ presents the heat of formation of the initial graphene fragment. The value $\Delta H_{nHgr}^{sk}$ can be obtained as a result of one-point-structure determination applied to the *n*-th equilibrium hydride after removing all hydrogen atoms. Attributed to the basal plane, $E_{def}^{tot}(n)$ has the form

$$E_{def}^{tot\ bs}(k) = E_{def}^{tot}(k+44) - E_{def}^{tot\ fr}(44). \tag{10}$$

Here $E_{def}^{tot\ fr}(44)$ presents the deformation energy of the framed membrane.

The deformation energy which accompanies each step of the hydrogenation can be determined as

$$E_{def}^{step\ bs}(k) = \Delta H_{(k+44)Hgr}^{sk} - \Delta H_{[(k+44)-1]Hgr}^{sk}, \tag{11}$$

where $\Delta H_{(k+44)Hgr}^{sk}$ and $\Delta H_{[(k=44)-1]Hgr}^{sk}$ match heats of formation of the carbon skeletons of the relevant hydrides at two subsequent steps of hydrogenation.

Similarly, the total and perstep chemical contributions caused by the formation of C-H bonds on the basal plane can be determined as

$$E_{cov}^{tot\ bs}(k) = E_{cpl}^{tot\ bs}(k) - E_{def}^{tot\ bs}(k) \tag{12}$$

and

$$E_{cov}^{step\ bs}(k) = E_{cov}^{tot}(k+44) - E_{cov}^{tot}(k+44-1). \tag{13}$$

Figure 7 displays the calculated total energies, attributed to the basal plane, for hydrides 1 and hydrides 2. The relevant perstep energies are shown in Fig.8. As seen in Fig.7, the total coupling energies $E_{cpl}^{tot\ bs}$ of both hydrides are negative by sign and gradually increase by absolute value in due course of increasing the number of adsorbed atoms. Besides, the absolute value growth related to hydrides 2 is evidently slow down starting at step 11 in contrast to the continuing growth for hydrides 1. This retardation is characteristic for other two energies presented in Fig.7 thus quantitatively distinguishing hydrides 2 from hydrides 1. The retardation of the growth of both $E_{cpl}^{tot\ bs}$ and $E_{cov}^{tot\ bs}$ energies obviously show that the addition of hydrogen to the fixed membrane of hydrides 2 at coverage higher than 30% is much more difficult than in the case of hydrides 1. This conclusion is supported by the behavior of perstep energies $E_{cpl}^{step\ bs}$ and $E_{cov}^{step\ bs}$ plotted in Fig.8a and 8c. If in the case of hydrides 1 the energy values oscillate around steady average values of -52 kcal/mol and -72 kcal/mol for $E_{cpl}^{step\ bs}$ and $E_{cov}^{step\ bs}$, respectively, in the case of hydrides 2, $E_{cpl}^{step\ bs}$ oscillates around average values that grow from -64 kcal/mol to -8 kcal/mol. Similar $E_{\text{cov}}^{step\ bs}$ oscillations occur around a general level that starts at -88 kcal/mod and terminates at -8 kcal/mol. Therefore, the reaction of the chemical attachment of hydrogen atoms for hydrides 1 is thermodynamically profitable through over covering that reaches 100% limit. In contrast, the large coverage for hydrides 2 becomes less and less profitable so that at final steps adsorption and desorption become competitive thus resulting in desorption of a hydrogen molecule, which was described in Section 4.

A particular attention should be given to changing the deformation of the carbon carcass caused by $sp^2 \rightarrow sp^3$ transformation of the electron configuration of the carbon atoms. Gradually increased by value for both hydride families, the energy $E_{def}^{tot\ bs}$ shown in Fig.7 describes strengthening the deformation in due course of growing coverage of the basal plane. Irregular dependence of $E_{def}^{step\ bs}$ on covering presented in Fig.8b allows for speaking about obvious topochemical character of each hydrogen attachment to the membrane basal plane. The topochemistry in this case implies chemical reactions occurred in a space subordinated to restricting conditions [52] like reactions on the solid surfaces considered earlier by Schmidt [53]. The disclosed dependences may serve as a direct manifestation of the reactions of such kind.

## 6. Discussion and conclusive remarks

The performed investigations have shown that both the hydrogenation of graphene itself and the final hydrides formed depend on several external factors, namely: 1) the state of the fixation of graphene substrate; 2) the accessibility of the substrate sides to the hydrogen; and 3) molecular or atomic composition of the hydrogen vapor. These circumstances make both computational consideration and technology of the graphene hydrogenation multimode with the number of variants not less than eight if only molecular and atomic adsorption do not occur simultaneously. A detailed consideration of all variants should be mandatory included in any serious project aimed at application of the hydrogen-graphene-based nanomaterials in general and for hydrogen-stored fuel cells, in particular. It is difficult to imagine how this program can be accomplished if not taking into account the odd electron correlation and not governing the hydrogenation process by the ACS algorithm. We have made only first steps on the way, performing investigation of the hydrogenation of the only (5, 5) nanographene. Nevertheless, the investigation has involved all the hydrogenation modes related to atomic adsorption and to two modes of the molecular adsorption. In all the cases, the hydrogen atoms attachment to the substrate carbon atoms was subordinated to the ACS algorithm described in Section 2. Taking together, the results allow for suggesting a rather integral picture of the events that accompany hydrogenation of graphene. It is summed up in Table 2. The importance of the obtained results is additionally supported by the fact that all the studies have been performed by using the same substrate that is the basal plane of the (5,5) nanographene with 44 framing hydrogen atoms. Additionally to the general picture, the following answers to crucial questions related to the hydrogenation of graphene can be suggested.

1. *Which kind of the hydrogen adsorption, namely, molecular or atomic, is the most probable?*

As follows from Table 2, our study has convincingly shown that only atomic adsorption is effective and energetically favorable, which is consistent with widely known fact of a practical absence of molecular hydrogen adsorption on graphite.

Figure 9 presents equilibrium structures of the graphene hydrides obtained in due course of the molecular adsorption of hydrogen on two-side accessible basal plane of free standing (a) and fixed (b) membranes. The coupling energy of the products was determined as

$$E_{cpl}^{tot}(n) = \Delta H_{nH_2gr} - \Delta H_{gr} - n\Delta H_{H_2}. \qquad (14)$$

Here $\Delta H_{nH_2gr}$, $\Delta H_{gr}$, and $\Delta H_{H_2}$ present heats of formation of the graphene hydride with $n$ hydrogen molecules, the pristine graphene, and the hydrogen molecule ($\Delta H_{H_2}$=-5.182 kcal/mol), respectively.

As seen in Fig.9a, the adsorption of the first hydrogen molecule on free standing membrane is accompanied with the formation of two C-H bonds of 1.18Å in length. The second molecule is not adsorbed at its approaching the substrate from either up or down side. However, keeping the molecule in the vicinity of the graphene substrate in the up position is followed by a significant lowering of the energy. Addition of the third molecule stimulates the adsorption of the second one while the third molecule is not adsorbed. As in the previous case, the formation of this complex leads to the energy gain. The forth molecule is willingly adsorbed while the fifth one as well as all the next up to the fifteenth is not adsorbed. Therefore, one may speak about 14% hydrogen covering of the basal plane.

In the case of fixed membrane (Fig. 9b), the adsorption of the first hydrogen molecule is endothermic and requires 11.7 kcal/mol. In contrast with the free standing membrane, none of the subsequent hydrogen molecules are adsorbed thus providing very low (4.5%) covering of the basal plane. At the same time, keeping the molecules in the vicinity of substrate is accompanied with the energy gain.

As for atomic adsorption, the formation of hydrides with practically total covering of the basal plane occurs quite possible. When both sides of the membrane are accessible to hydrogen atoms, the hydrogenation of graphene is completed by the formation of the 100% covered regularly structured graphane (Fig.10a) whose description is given in Section 3. If the membrane is accessible from only one side, the consequent attachment of hydrogen atoms to the substrate causes arching of its carbon skeleton that takes the shape of a canopy at the final stage (Fig.10b). In both cases, C-C bonds lengthen taking 1.51-1.53Å. However, under fixation of the membrane edges, not all the bonds are able to meet the requirement so that a part of them should stay quite short. Under this condition, a pair of hydrogen atoms which had to be attached to two carbon atoms forming the bonds is not allowed to perform the job thus stimulating atoms to associate and to form a hydrogen molecule outside of the basal plane. In the current study, the last atoms 43 and 44 had such fate, which resulted in desorption of one hydrogen molecule and lowering the plane covering up to 96%.

At first glance, the final stage of the two-side adsorption of hydrogen atoms on the free standing membrane (Fig.10c) seems to be similar to that presented in Fig.10a. However, attentive consideration reveals that the chairlike configuration of cyclohexanoids, regularly composed in the upper side of the sample, is violated when approaching the sample bottom so that we have to speak about mixing of conformers that causes the distortion of the regular structure of the carbon skeleton. This circumstance does not prevent from achieving 100% filling of the basal plane. But the sample itself presents a mixture of a regular graphane area neighbouring with some elements of amorphous structure. Obviously, the partial contribution of each component depends on the graphene sample size.

In contrast to the above said, the one-side adsorption of the free standing membrane has resulted in the formation of a peculiar basket that is formed when two ends of a rectangular figure situated over its diagonal are closely approached to each other (Fig.10d). The formation of the 100% hydride of so peculiar shape is accompanied by the energy gain of ~ 1kcal/mol per each carbon atom.

The reason for so dramatic difference between atomic and molecular adsorption might be a consequence of the tendency of the graphene substrate to conserve the hexagon pattern. But obviously, the pattern conservation can be achieved if only the substrate hydrogenation provides the creation of cyclohexane-patterned structure that corresponds to either one of the conformers of the latter, or their mixture. If non-coordinated deposition of individual atoms, as we saw in Fig. 3 and Fig. 4, can meet the requirement, a coordinated deposition of two atoms on neighboring carbons of substrate evidently make the formation of a cyclohexane-conformer pattern much less probable thus making molecular adsorption unfavorable.

*2. What is characteristic image of the hydrogen atom attachment to the substrate?*

The hydrogen atom is deposited on-top of the carbon ones in both up and down configurations. In contrast to a vast number of organic molecules, the length of C-H bonds formed under adsorption exceed 1.10Å, therewith differently for different adsorption events. Thus, C-H bonds are quite constant by value of 1.122Å in average for framing hydrogens that saturate edge carbon atoms of the substrate. Deposition on the basal plane causes enlarging the value up to maximum 1.152Å [53]. However, the formation of a regular chairlike cyclohexane structure like graphane leads to equalizing and shortening the bonds to 1.126Å. The above picture, which is characteristic for the fixed membrane, is significantly violated when going to one-side deposited fixed membrane or two-side deposited free standing membrane that exhibits the difference in the strength of the hydrogen atoms coupling with the related substrates.

*3. Which carbon atom is the first target subjected to the hydrogen attachment?*
And
*4. How carbon atoms are selected for the next steps of the adsorption?*

Similarly to fullerenes and carbon nanotubes [33], the formation of graphene polyhydrides $(CH)_n$ has been considered in the framework of algorithmic stepwise computational synthesis, each subsequent step of which is controlled by the distribution of atomic chemical susceptibility in terms of partial numbers of effectively unpaired electrons on atom, $N_{DA}$, of preceding derivative over the substrate atoms. The quantity is a direct consequence of the odd electron correlation. The high-rank $N_{DA}$ values definitely distinguish the atoms that should serve as targets for a forthcoming chemical attack. Additionally, the LTE criterion has provided the choice of the most energetically stable hydride. The successful generation of the polyderivative families of fluorides [54] and hydrides [55] as well as other polyderivatives of fullerene $C_{60}$ [33], of 100% polyhydride $(CH)_n$ related to chairlike graphane described in the current paper as well as 96% of tablelike-cyclohexane hydride $(CH)_n$ of the canopy shape has shown a high efficacy of the approach in viewing the process of the polyderivatives formation which makes it possible to proceed with a deep insight into the mechanism of the chemical modification.

5. *Is there any connection between the sequential adsorption pattern and cyclohexane conformers formed in due course of hydrogenation?*

The performed investigations have shown that there is a direct connection between the state of graphene substrate and the conformer pattern of the polyhydride formed. The pattern is governed by the cyclohexane conformer whose formation under ambient conditions is the most profitable. Thus, a regular chairlike-cyclohexane conformed graphene with 100% hydrogen covering, known as graphane, is formed in the case when the graphene substrate is a perimeter-fixed membrane, both sides of which are accessible for hydrogen atoms. When the membrane is two-side accessible but its edges are not fixed, the formation of a mixture of chairlike and boatlike cyclohexane patterns has turned out more profitable. As shown in the current paper, the polyhydride total energy involves deformational and covalent components. That is why the difference in the conformer energy in favor of chairlike conformer formed in free standing membrane is compensated by the gain in the deformation energy of the carbon carcass caused by the formation of boatlike conformer, which simulates a significant corrugation of the initial graphene plane. The mixture of the two conformers transforms therewith a regular crystalline behavior of graphane into a partially amorphous-like behavior in the latter case.

When the fixed membrane is one-side accessible, the configuration produced is rather regular and looks like an infinite array of *trans*-linked tablelike cyclohexane conformers. The coupling of hydrogen atoms with the carbon skeleton is the weakest among all the considered configurations, which is particularly characterized by the longest C-H bonds of 1.18-1.21Å in length. The carbon skeleton takes a shape of a canopy exterior. Energetically rich conditions similar, for example, to hydrothermal ones will obviously favor the hydrogenation of this kind. This finding allows shredding light onto the genesis of mineralloid shungit [56], consisting of globules of compressed concave-saucer-like graphene sheets, whose formation occurs under high temperature and pressure as well as under high concentration of hydrogen atoms.

ACKNOWLEDGEMENTS

The authors greatly appreciate fruitful discussions with K. Novoselov and Yu.S.Nechaev

**REFERENCES**

1. Nechaev YuS: The high-density hydrogen carrier intercalation in graphane-like nanostructures. Relevance to its on-board storage in fuel-cell-powered vehicles. *The Open Fuel Cells Journal,* 2011, 4: 16-29.
2. Nechaev YuS: On the solid hydrogen carrier intercalation in graphane-like regions in carbon-based nanostructures. *International Journal of Hydrogen Energy,* 2011, 36: 9023-9031.
3. Elias DC, Nair RR, Mohiuddin TMG, Morozov SV, Blake P, Halsall MP, Ferrari AC, Boukhvalov DW, Katsnelson MI, Geim AK, Novoselov KS: Control of graphene's properties by reversible hydrogenation: evidence for graphane. *Science*, 2009, 323: 610-613.
4. Sofo JO, Chaudhari AS, Barber GD: Graphane: a two-dimensional hydrocarbon. *Phys. Rev. B*, 2007, 75: 153401.
5. Balog R, Jørgensen B, Nilsson L, Andersen M, Rienks E, Bianchi M, Fanetti M, Lægsgaard E, Baraldi A, Lizzit S, Sljivancanin Z, Besenbacher F, Hammer B, Pedersen

TG, Hofmann P, Hornekær L: Bandgap opening in graphene induced by patterned hydrogen adsorption. *Nature Materials,* 2010, 9: 315- 319.
6. Lee J- H, Grossman JC: Magnetic properties in graphene-graphane superlattices. *Appl. Phys. Lett.*, 2010, 97: 133102.
7. Zhou J, Wang Q, Sun Q, Chen XS, Kawazoe Y, Jena P: Ferromagnetism in semihydrogenated graphene sheet. *Nano Lett.*, 2009, 9: 3867-3870.
8. Savini G, Ferrari AC, Giustino F: First-principles prediction of doped graphane as a high-temperature electron-phonon superconductor. *Phys. Rev. Lett.*, 2010, 105: 037002.
9. Cudazzo P, Attaccalite C, Tokatly IV, Rubio A: Strong charge-transfer excitonic effects and Bose-Einstein exciton-condensate in graphane. *Phys. Rev. Lett.*, 2010, 104*:* 226804.
10. Crassee I, Levallois J, Walter AL, Ostler M, Bostwick A, Rotenberg E, Seyller T, van der Marel D, Kuzmenko AB: Giant Faraday rotation in single- and multilayer graphene. *Nature Physics*, 2011, 7: 48-51.
11. Singh AK, Penev ES, Yakobson BI: Vacancy clusters in graphane as quantum dots. *ACS Nano*, 2010, 4: 3510-3514.
12. Muñoz E, Singh AK, Ribas MA, Penev ES, Yakobson BI: The ultimate diamond slab: graphAne versus graphEne. *Diam. Rel. Mat.*, 2010, 19: 368-373.
13. Topsakal M, Cahangirov S, Ciraci S: The response of mechanical and electronic properties of graphane to the elastic strain. *Appl. Phys. Lett.*, 2010, 96: 091912.
14. Pei QX, Zhang YW, Shenoy VB: A molecular dynamics study of the mechanical properties of hydrogen functionalized graphene. *Carbon*, 2010, 48: 898-904.
15. Openov LA, Podlivaev AI: Spontaneous regeneration of an atomically sharp graphene/graphane interface under thermal disordering. *JETP Lett.*, 2010, 90: 459-463.
16. Openov LA, Podlivaev AI: Thermal desorption of hydrogen from graphane. *Tech. Phys. Lett.*, 2010, 36: 31-33.
17. Schmidt MJ, Loss D: Edge states and enhanced spin-orbit interaction at graphene/graphane interfaces. *Phys. Rev. B,* 2010, 81: 165439.
18. AlZahrania AZ, Srivastava GP: Structural and electronic properties of H-passivated graphene. *Appl. Surf. Sci.*, 2010, 256: 5783-5788.
19. Leenaerts O, Peelaers H, Hernandez-Nieves AD, Partoens B, Peeters FM: First-principles investigation of graphene fluoride and graphane. Phys. Rev. B, 2010, 82: 195436.
20. Bhattacharya A, Bhattacharya S, Majumder C, Das GP: The third conformer of graphane: A first principles DFT based study. *Phys. Rev. B*, 2010*,* 83: 033404.
21. Sahin H, Ataca C, Ciraci S: Electronic and magnetic properties of graphane nanoribbons. *Phys. Rev. B*, 2010, 81: 205417.
22. Ju W, Wang H, Li T, Fu Zh, Zhang Q: First-principals study on the three-dimensional structure of graphane. *Key Engineer. Mat.*, 2010, 434-435: 803-804.
23. Rosas JJH, Gutiérrez RER, Escobedo-Morales A, Chigo Anota E: First principles calculations of the electronic and chemical properties of graphene, graphane, and graphene oxide. *J. Mol. Model.*, 2011, 17: 1133-1139.
24. Sheka EF, Popova NA: Mechanochemical reaction in graphane under uniaxial tension. *J. Phys. Chem. C,* 2011, 115: 23745–23754.
25. Allouche A, Jelea A, Marinelli F, Ferro Y: Hydrogenation and dehydrogenation of graphite (0001) surface: a density functional theory study. *Phys. Scr.*, 2006, 124: 91-95.
26. Jeloaica L, Sidis V: DFT investigation of the adsorption of atomic hydrogen on a cluster-model graphite surface. *Chem. Phys. Lett.*, 1999, 300: 157-163.
27. Sha X, Jackson B: First-principles study of the structural and energetic properties of H atoms on a graphite (0 0 0 1) surface. *Surf. Sci.*, 2002, 496: 318-330.

28. Hornekær L, Sljivancanin ZS, Xu W, Otero R, Rauls E, Stensgaard I, Lægsgaard E, Hammer B, Besenbacher F: Metastable structures and recombination pathways for atomic hydrogen on the graphite (0001) surface. *Phys. Rev. Lett.*, 2006, 96: 156104.
29. Ito A, Nakamure H, Takayama A: Chemical reaction between single hydrogen atom and graphene. *arXiv* 0703377v2 [cond-mat.other], 2007.
30. Casolo S, Lovvik OM, Martinazzo R, Tantardini GF: Understanding adsorption of hydrogen atoms on graphene. *J. Chem. Phys.*, 2009, 130: 054704.
31. Flores MZS, Autreto PAS, Legoas SB, Galvao S: Graphene to graphane: a theoretical study. *Nanotechnology*, 2009, 20: 465704.
32. Davidson ER and Clark AE: Analysis of wave functions for open-shell molecules. *Phys Chem. Chem. Phys.*, 2007. 9: 1881-94.
33. Sheka EF: *Fullerene Nanoscience : Nanochemistry, Nanomedicine, Nanophotonics, Nanomagnetism.* Boca Raton: CRC Press, Taylor and Francis Group; 2011.
34. Davidson E: How robust is present-day DFT? *Int. J. Quant. Chem.* 1998, 69: 214-45.
35. Kaplan I: Problems in DFT with the total spin and degenerate states. *Int. J. Quant. Chem.*, 2007, 107: 2595-603.
36. Takatsuka K, Fueno T, Yamaguchi K: Distribution of odd electrons in ground-state molecules. *Theor. Chim. Acta,* 1978, 48: 175-83.
37. Staroverov VN, Davidson ER: Distribution of effectively unpaired electrons. *Chem. Phys. Lett.* 2000, 330: 161-168.
38. Benard M: A study of Hartree-Fock instabilities in $Cr_2(O_2CH)_4$ and $Mo_2(O_2CH)_4$. *J.Chem. Phys.* 1979, 71: 2546-56.
39. Lain L, Torre A, Alcoba DR, Bochicchio RC: A study of the relationships between unpaired electron density, spin-density and cumulant matrices. *Theor. Chem. Acc.* 2011, 128:405–410.
40. Sheka EF, Chernozatonskii LA: Bond length effect on odd electrons behavior in single-walled carbon nanotubes. *J. Phys. Chem. C*, 2007, 111: 10771-10780.
41. Gao X, Zhou Z, Zhao Y, Nagase S, Zhang SB, Chen Z: Comparative study of carbon and BN nanographenes: Ground electronic states and energy gap engineering. *J. Phys. Chem. A*, 2008, 112: 12677-12682.
42. Zayets VA: CLUSTER-Z1: Quantum-Chemical Software for Calculations in the s,p-Basis. Kiev: Institute of Surface Chemistry Nat. Ac.Sci. of Ukraine. 1990.
43. Sheka EF, Chernozatonskii LA: Broken spin symmetry approach to chemical susceptibility and magnetism of graphenium species. *J. Exp. Theor. Phys.*, 2010, 110: 121-132.
44. Sheka EF, Chernozatonskii LA: Chemical reactivity and magnetism of graphene. *Int. Journ. Quant. Chem.*, 2010, 110: 1938-1946.
45. Sheka EF, Chernozatonskii LA: Odd-electron approach to covalent chemistry and magnetism of single-walled carbon nanotubes and graphene. *Nanostruct. Mat. Phys. Model.*, 2009, 1: 115-149.
46. Sheka EF, Zayets VA: The radical nature of fullerene and its chemical activity. *Russ. J. Phys. Chem.* 2005, 79: 2009-2014.
47. Sheka EF: Chemical susceptibility of fullerenes in view of Hartree-Fock approach. *Int. Journ. Quant. Chem.*, 2007, 107: 2803-16.
48. Berzigiyarov PK, Zayets VA, Ginzburg IYa, Razumov VF, Sheka EF: NANOPACK: Parallel codes for semiempirical quantum chemical calculations of large systems in the *sp*- and *spd*-basis. *Int. J. Quant. Chem.* 2002, 88:449-462.
49. Meyer JC, Girit CO, Crommie MF, Zettl A: Imaging and dynamics of light atoms and molecules on graphene. Nature, 2008, 454: 319.

50. Sheka EF, Popova NA : When a covalent bond is broken? *arXiv*:1111.1530v1 [physics.chem-ph]. 2011.
51. Sheka EF, Shaymardanova LKh: $C_{60}$-based composite in view of topochemical reaction. *J. Mat. Chem.*, 2011, 21, 17128 – 17146.
52. Schmidt GMJ: Photodimerization in the solid state. *Pure Appl. Chem.* 1971, 27: 647-678.
53. Sheka EF, Popova NA: Odd-electron molecular theory of the graphene hydrogenation. *J. Mol. Mod.*, 2012 (in press).
54. Sheka EF: Step-wise computational synthesis of fullerene $C_{60}$ derivatives. Fluorinated fullerenes $C_{60}F_{2k}$. *J. Exp. Theor. Phys.*, 2010, 111: 395-412.
55. Sheka EF: Computational synthesis of hydrogenated fullerenes from $C_{60}$ to $C_{60}H_{60}$. *J. Mol. Mod.*, 2011, 17: 1973-1984.
56. Sheka EF, Rozhkova NP, Popova NA: private communication: 2011.

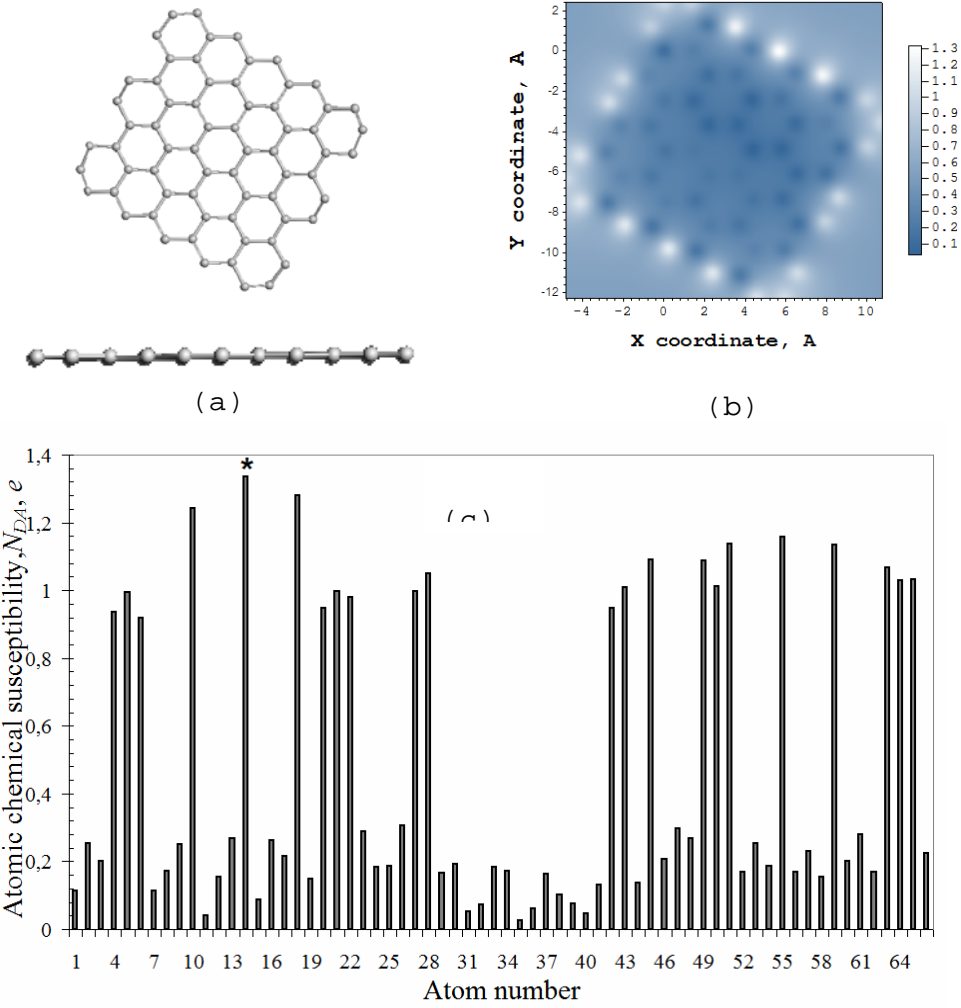

**Figure 1**. Top and side views of the equilibrium structure of (5,5) nanographene (a) and ACS distribution over atoms in real space (b) and according to atom numbers in the output file (c).

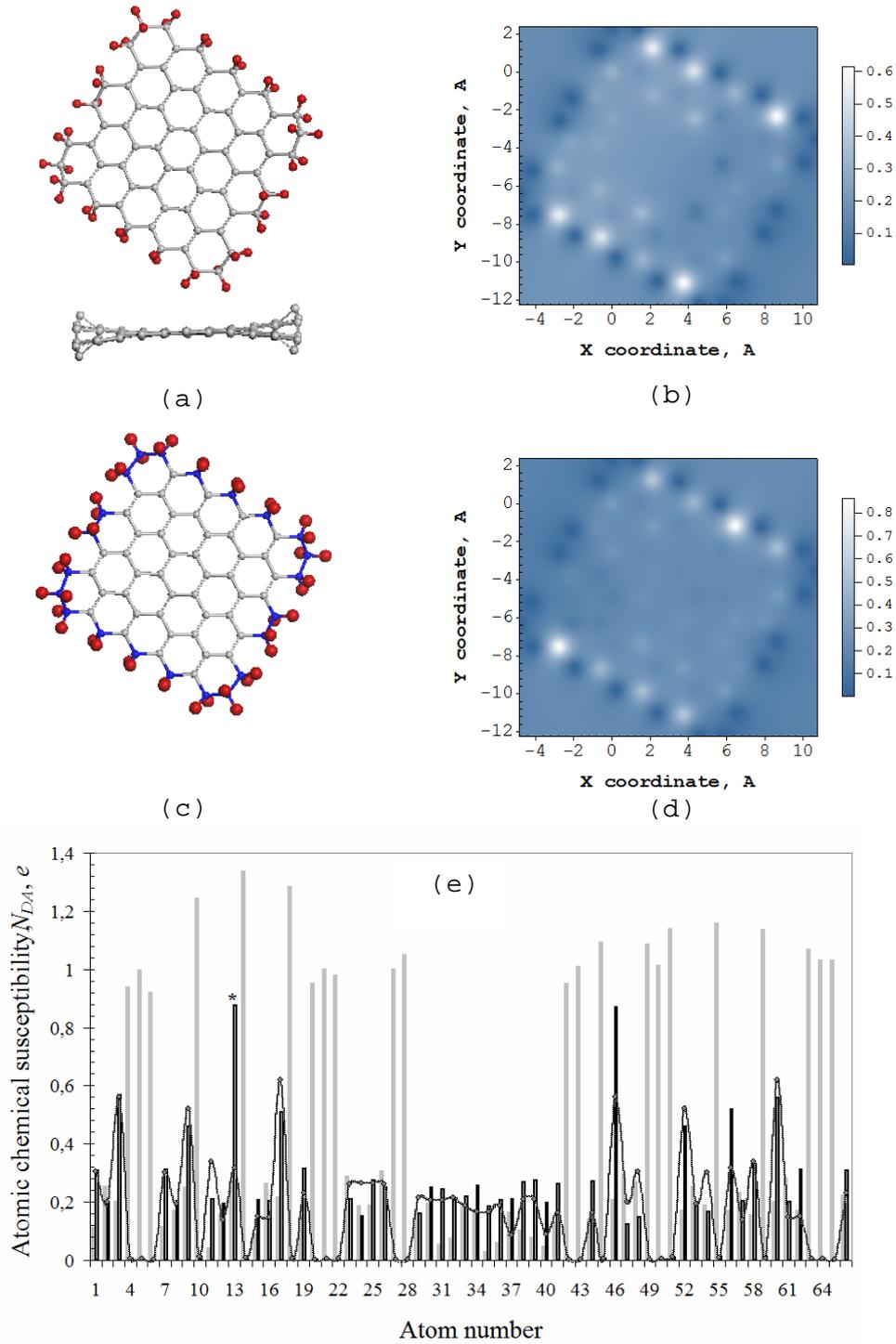

**Figure 2**. Equilibrium structures of free standing (top and side views) (a) and fixed (c) (5,5) graphene membrane and ACS distribution over atoms in real space (b, d) and according to the atom numbers in output file (e). Light gray histogram plots ACS data for the pristine (5,5) nanographene. Curve and black histogram are related to membranes in panels *a* and *c*, respectively.

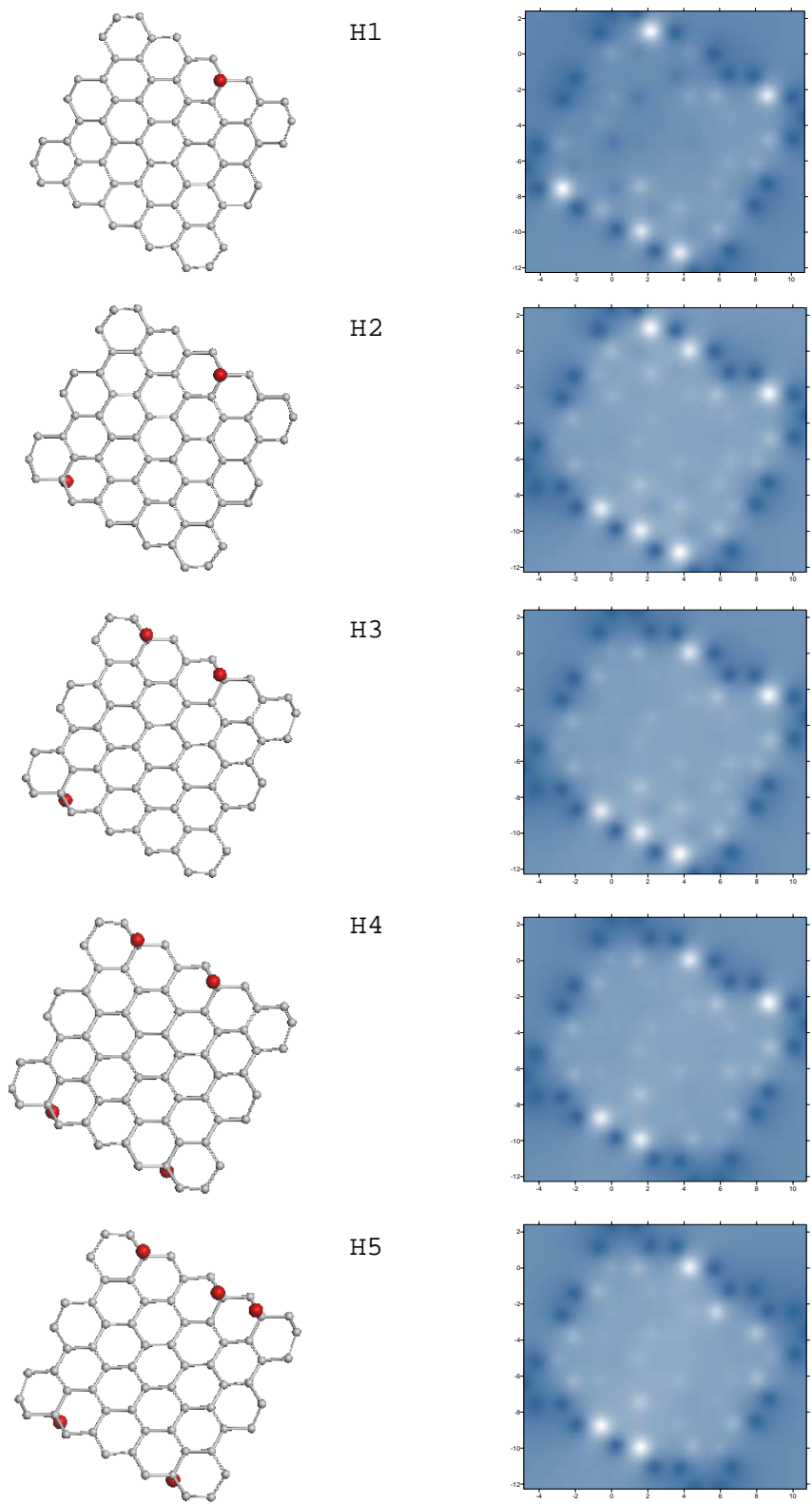

Figure 3 contnd

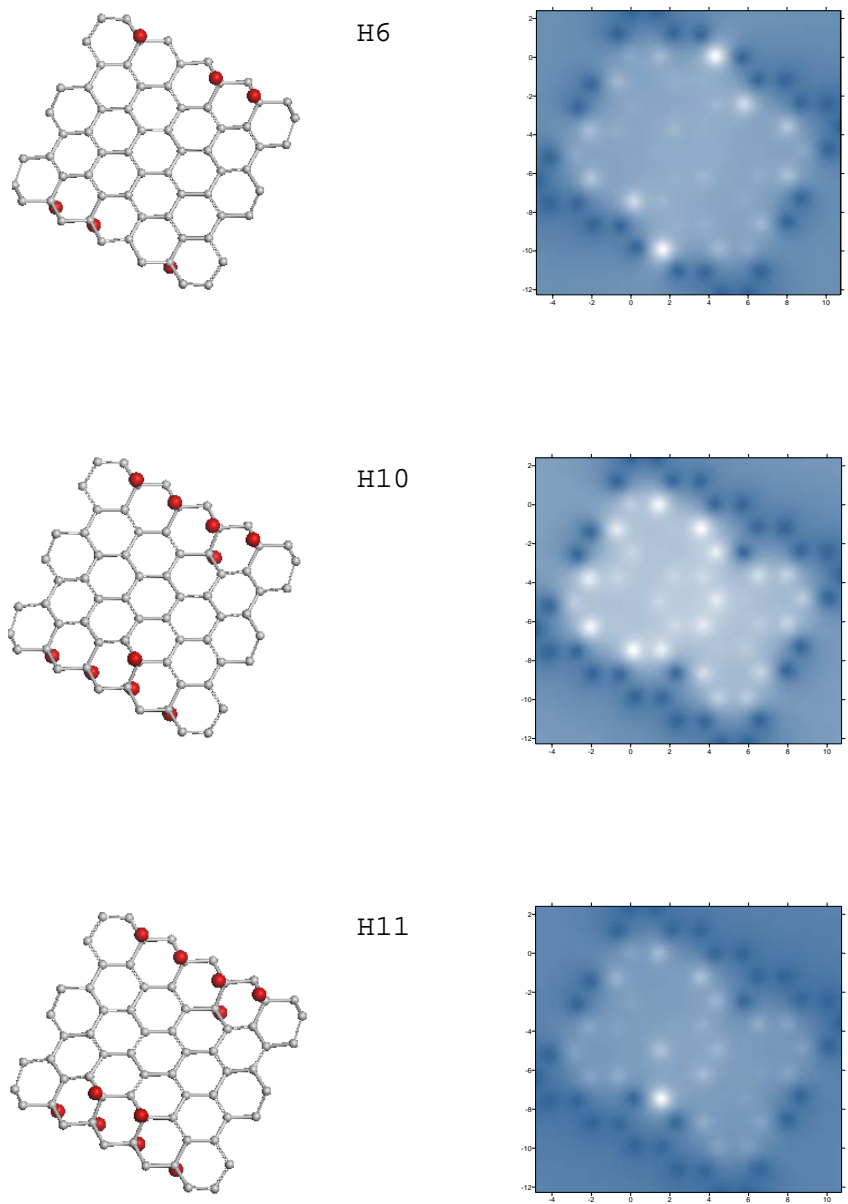

**Figure 3.** Equilibrium structures (left) and real-space ACS maps (right) of hydrides 1 related to initial stage of the basal-plane hydrogenation. HKs denote hydrides with K hydrogen atoms deposited on the membrane.

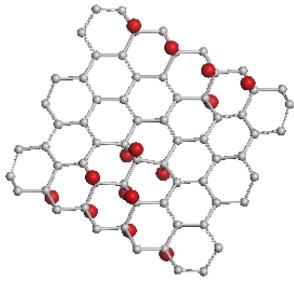 H15 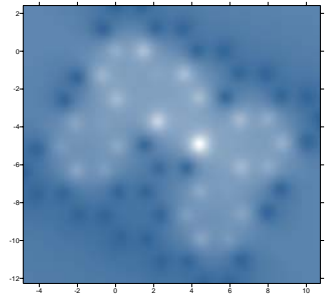

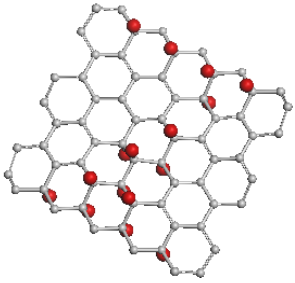 H16 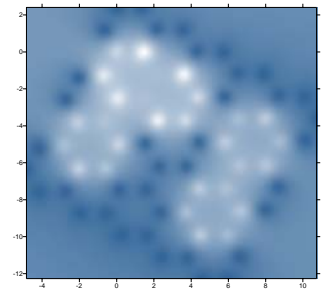

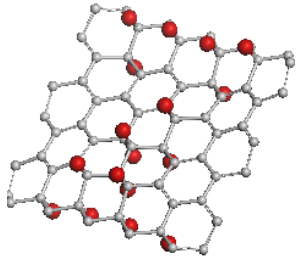 H17 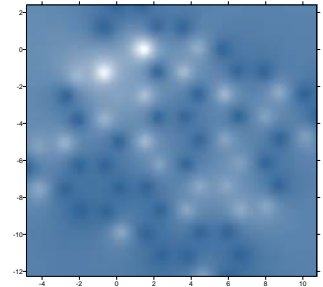

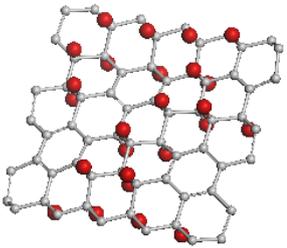 H25 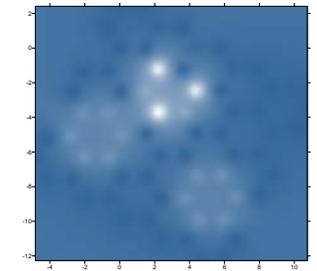

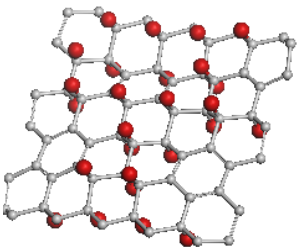 H26 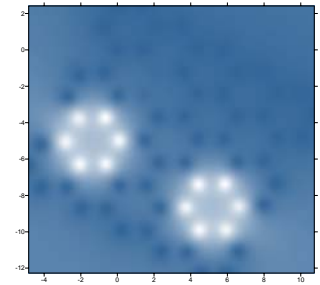

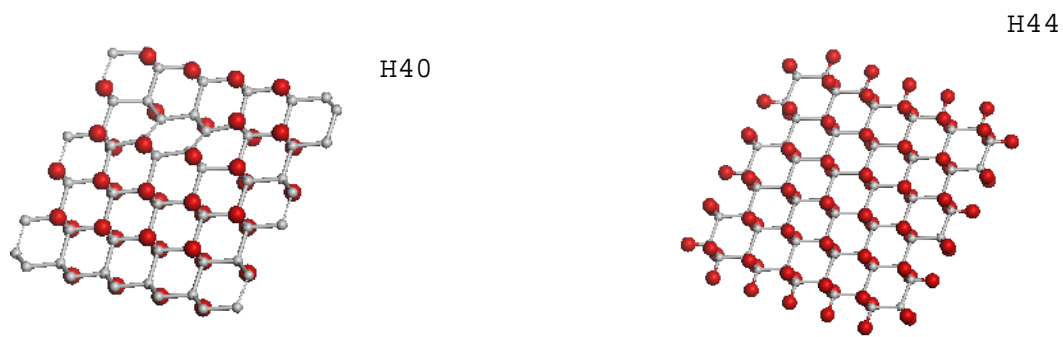

**Figure 4.** Equilibrium structures (left) and real-space ACS maps (right) of hydrides related to the middle and conclusive stages of the basal-plane hydrogenation. See caption to Fig.3.

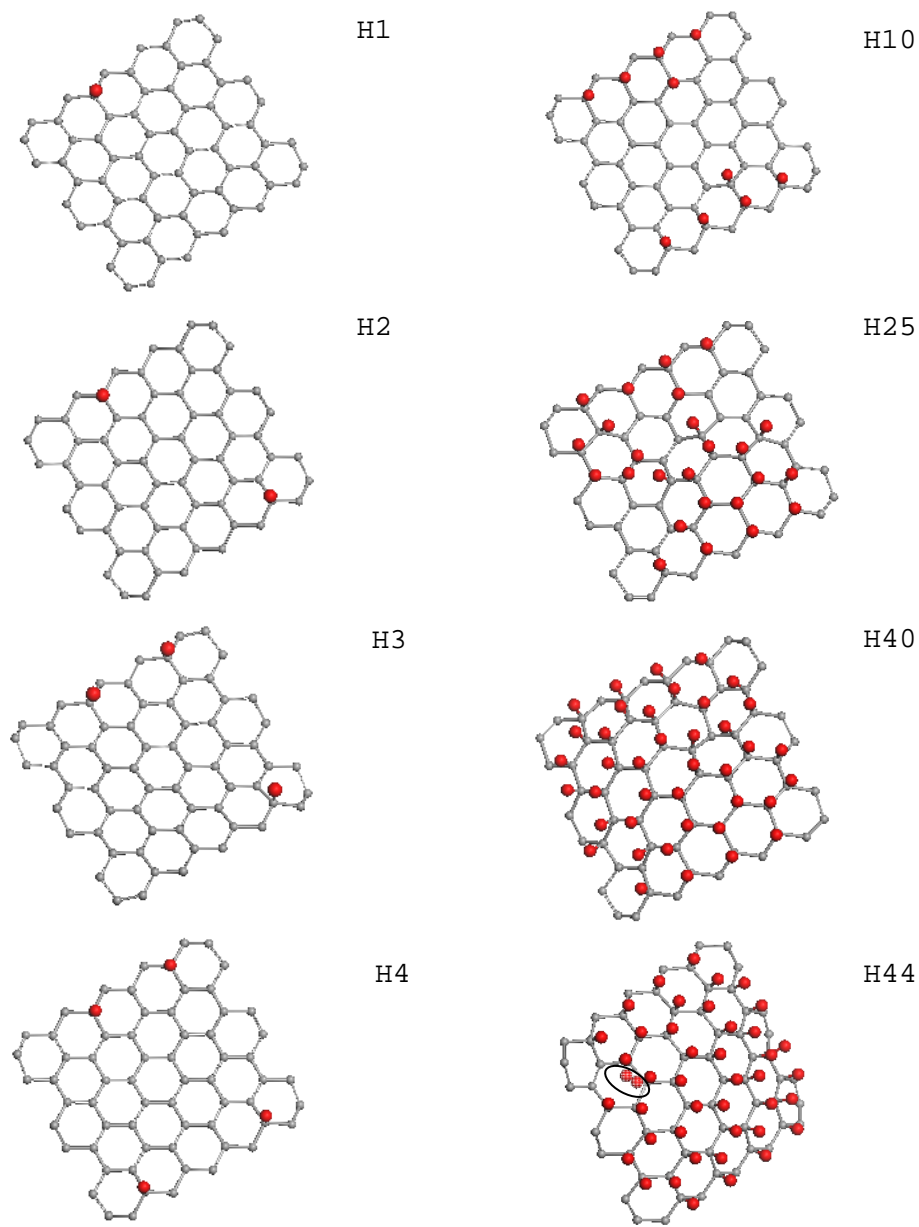

**Figure 5.** Equilibrium structures of hydrides 2 formed at the one-side basal-plane hydrogenation of the fixed (5,5) graphene membrane. See caption to Fig.3. The desorbed hydrogen molecule from hydride 44 is marked by ellipse (see text).

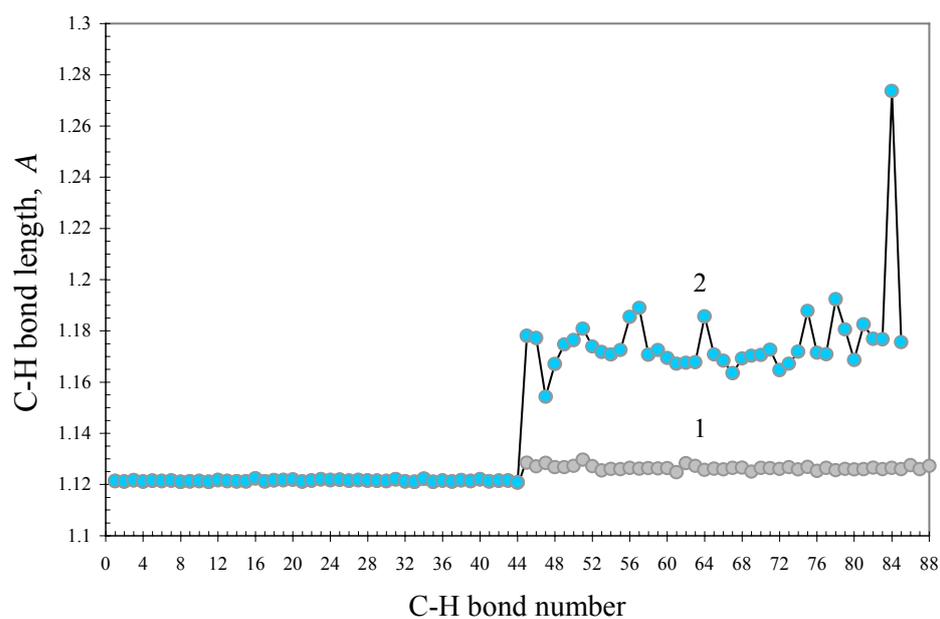

**Figure 6**. C-H bond length distribution for H44 hydrides of families 1 (1) and 2 (2).

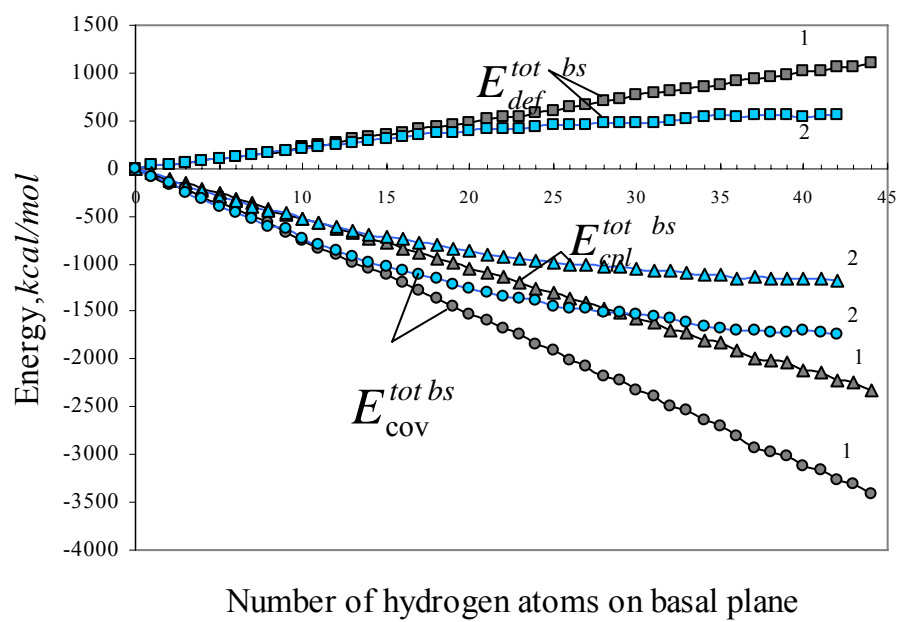

**Figure 7.** Total energies of coupling $E_{cpl}^{tot\ bs}$ (triangles), deformation $E_{def}^{tot\ bs}$ (squares) and covalent bonding $E_{cov}^{tot\ bs}$ (circles) via the number of hydrogen atoms deposited on the basal plane for hydrides 1 (dark gray) and 2 (light blue) according to Exs. (7), (10), and (12), respectively.

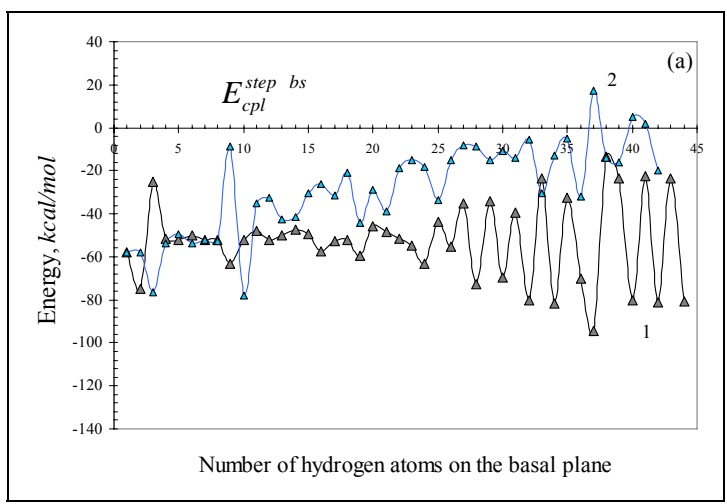

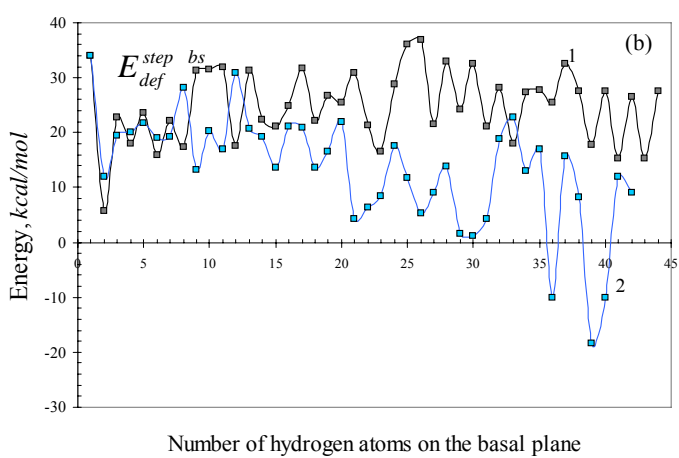

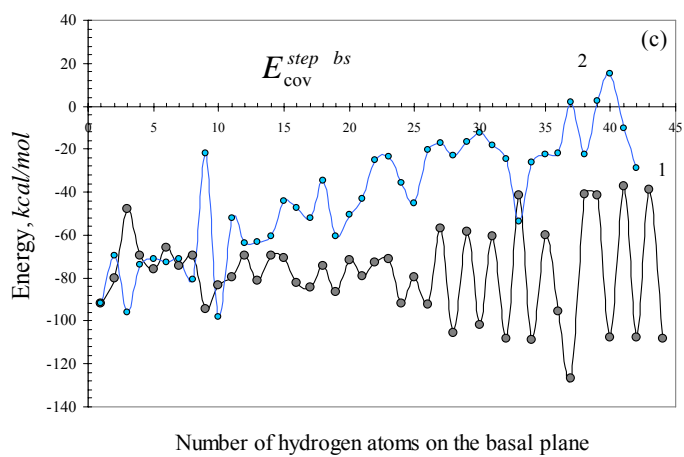

**Figure 8.** Perstep energies of coupling $E_{cpl}^{step\ bs}$ (a), deformation $E_{def}^{step\ bs}$ (b) and covalent bonding $E_{cov}^{step\ bs}$ (c) via the number of hydrogen atoms deposited on the basal plane for hydrides 1 (dark gray) and 2 (light blue) according to Exs. (8), (11), and (13), respectively.

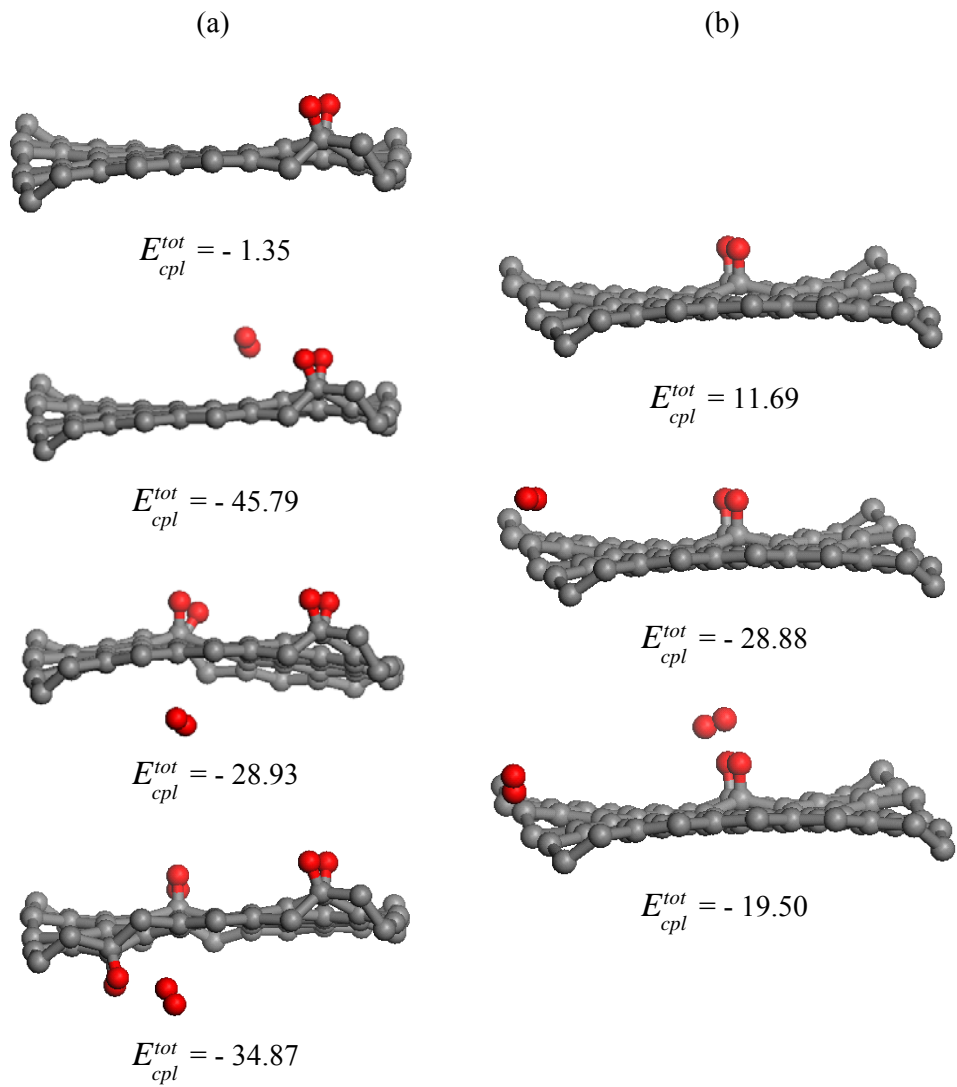

**Figure 9**. Equilibrium structure of hydrides formed at the molecular adsorption of hydrogen on the free standing (a) and fixed (b) (5,5) graphene membranes, accessible to the adsorbate from both side. $E_{cpl}^{tot}$ is the total coupling energy (see (14)). The framing hydrogen atoms are not shown.

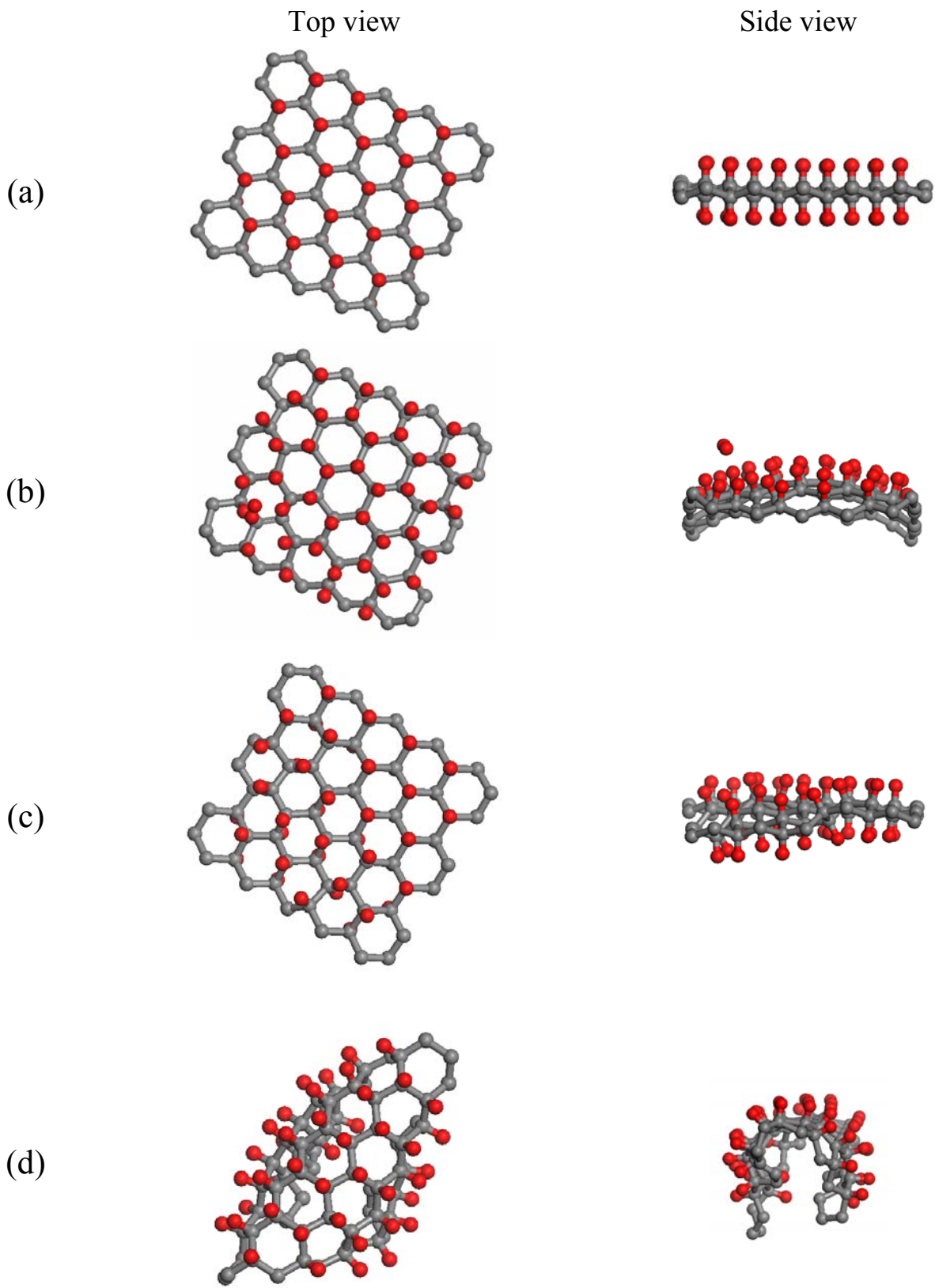

**Figure 10**. Top and side views of the equilibrium structures of hydrides formed at the atomic adsorption of hydrogen on the fixed (a, b) and free standing (c, d) (5,5) graphene membrane, accessible to the adsorbate from both (a, c) and one (b, d) sides.

**Chart 1.** Explication of the subsequent steps of the hydrogenation of (5, 5) nanographene. $N_{at}$ number carbon atoms, $N_{DA}$ presents the atomic chemical susceptibility of the atoms, ΔH is the energy of formation of the hydrides in kcal/mol.

| H0 | | H1 (13) | | H2 (46) | | H3 (3) | | H4 (60) | | H5 (17) | | H6 (52) | | H7 (9) | |
|---|---|---|---|---|---|---|---|---|---|---|---|---|---|---|---|
| $N_{at}$ | $N_{DA}$ | $N_{at}$ | $N_{DA}$ | $N_{at}$ | $N_{DA}$ | $N_{at}$ | $N_{DA}$ | $N_{at}$ | $N_{DA}$ | $N_{at}$ | $N_{DA}$ | $N_{at}$ | $N_{DA}$ | $N_{at}$ | $N_{DA}$ |
| 13 | 0,87475 | | | | | | | | | | | | | | |
| 46 | 0,87034 | up | | up | | up | | up | | up | | up | | up | | 
| 3 | 0,56633 | 46 | 0,62295 | 3 | 0,56119 | 60 | 0,55935 | 17 | 0,54497 | 52 | 0,49866 | 56 | 0,47308 | 56 | 0,47318 |
| 60 | 0,55862 | 3 | 0,61981 | 60 | 0,55965 | 17 | 0,54302 | 56 | 0,50027 | 56 | 0,49856 | 9 | 0,47072 | 12 | 0,42593 |
| 56 | 0,51946 | 60 | 0,56055 | 17 | 0,54299 | 56 | 0,51849 | 52 | 0,49996 | 9 | 0,47193 | 12 | 0,36783 | 8 | 0,42263 |
| 17 | 0,50841 | ΔH = 126.50 | | ΔH = 101.88 | | ΔH = 96.88 | | ΔH = 99.77 | | ΔH = 95.00 | | ΔH = 96.82 | | ΔH = 97.50 | |
| 52 | 0,45883 | down | | down | | down | | down | | down | | down | | down | |
| 9 | 0,45843 | 46 | 0,62197 | 3 | 0,56174 | 60 | 0,55906 | 17 | 0,54576 | 52 | 0,49684 | 9 | 0,47392 | 56 | 0,46883 |
| 58 | 0,33071 | 3 | 0,62099 | 60 | 0,55926 | 17 | 0,54279 | 56 | 0,49878 | 56 | 0,49602 | 56 | 0,47247 | 12 | 0,41783 |
| 19 | 0,31539 | 60 | 0,55979 | 17 | 0,54297 | 56 | 0,51736 | 52 | 0,49782 | 9 | 0,47141 | 12 | 0,36857 | 8 | 0,41459 |
| | | ΔH = 127.54 | | ΔH = 98.19 | | ΔH = 101.90 | | ΔH = 95.99 | | ΔH = 99.40 | | ΔH = 96.23 | | ΔH = 98.92 | |

| H8 (56) | | H9 (12) | | H10 (57) | | H11 (53) | | H12 (58) | | H13 (33) | | H14 (30) | | H15 (34) | |
|---|---|---|---|---|---|---|---|---|---|---|---|---|---|---|---|
| $N_{at}$ | $N_{DA}$ | $N_{at}$ | $N_{DA}$ | $N_{at}$ | $N_{DA}$ | $N_{at}$ | $N_{DA}$ | $N_{at}$ | $N_{DA}$ | $N_{at}$ | $N_{DA}$ | $N_{at}$ | $N_{DA}$ | $N_{at}$ | $N_{DA}$ |
| up | | up | | up | | up | | up | | up | | up | | up | |
| 12 | 0,42462 | 57 | 0,42824 | 53 | 0,30873 | 58 | 0,54014 | 33 | 0,33339 | 30 | 0,40225 | 34 | 0,34848 | 35 | 0,51203 |
| 8 | 0,42404 | 53 | 0,42131 | 8 | 0,30851 | 30 | 0,32126 | 8 | 0,30396 | 34 | 0,3618 | 31 | 0,33662 | 31 | 0,36904 |
| 53 | 0,41554 | 47 | 0,30877 | 2 | 0,2956 | 8 | 0,32075 | 2 | 0,29194 | 8 | 0,34811 | 8 | 0,2934 | 2 | 0,29541 |
| ΔH = 99.83 | | ΔH = 118.04 | | ΔH = 79.46 | | ΔH = 74.17 | | ΔH = 73.81 | | ΔH = 65.87 | | ΔH = 70.90 | | ΔH = 73.11 | |
| down | | down | | down | | down | | down | | down | | down | | down | |
| 12 | 0,4232 | 57 | 0,42759 | 8 | 0,30801 | 58 | 0,58671 | 33 | 0,32811 | 30 | 0,40175 | 34 | 0,34603 | 35 | 0,5002 |
| 57 | 0,42303 | 53 | 0,4216 | 2 | 0,29368 | 8 | 0,32092 | 8 | 0,30638 | 34 | 0,3625 | 31 | 0,3272 | 31 | 0,36752 |
| 8 | 0,42294 | 47 | 0,31082 | 53 | 0,29305 | 30 | 0,30986 | 2 | 0,28938 | 8 | 0,34298 | 2 | 0,29516 | 2 | 0,29455 |
| ΔH = 98.98 | | ΔH = 89.19 | | ΔH = 107.96 | | ΔH = 101.20 | | ΔH = 68.89 | | ΔH = 84.89 | | ΔH = 67.58 | | ΔH = 71.83 | |

**Table 1**. Identifying parameters of the odd electrons correlation in graphene fragments

| Fragment $(n_a, n_z)$ | Odd electrons $N_{odd}$ | $\Delta E^{RU}$ | | $N_D$ | | $\hat{S}_U^2$ |
|---|---|---|---|---|---|---|
| | | kcal/mol | $\delta E^{RU} = \Delta E^{RU} / E^R$ % | $e^-$ | $\delta N_D = N_D / N_{odd}$ % | |
| (5, 5) | 88 | 307.6 | 17 | 31 | 35 | 15.5 |
| (7, 7) | 150 | 376.2 | 15 | 52.6 | 35 | 26.3 |
| (9, 9) | 228 | 641.6 | 19 | 76.2 | 35 | 38.1 |
| (11, 10) | 296 | 760.1 | 19 | 94.5 | 32 | 47.2 |
| (11, 12) | 346 | 900.6 | 20 | 107.4 | 31 | 53.4 |
| (15, 12) | 456 | 1038.1 | 19 | 139 | 31 | 69.5 |

**Table 2.** Final products of the hydrogen adsorption on (5,5) nanographene

| *Atomic adsorption* | Two-side access | One-side access | *Molecular adsorption* | Two-side access | One-side access |
|---|---|---|---|---|---|
| Fixed membrane | 100%-covered hydride of crystalline graphane structure (Fig. 8a) | 96%-covered hydride of canopy structure (Fig. 8b) | Fixed membrane | ~5%-covered hydride of non-planar sheet structure (Fig. 7b) | No results |
| Free standing membrane | 100%-covered hydride of crystalline graphane structure accompanied by fragments of amorphous structure (Fig. 9a) | 100%-covered hydride of a basket structure (see text) (Fig. 9b) | Free standing membrane | 10-15%-covered hydride of irregular structure of bent sheet (Fig. 7a) | No results |